\documentclass[aip,rbi,reprint,graphicx]{revtex4-1} % for checking your page length
\usepackage[dvips]{graphicx}
\usepackage{color}
\begin{document}

%\preprint{}

\title{Radio-transparent multi-layer insulation for radiowave receivers} 
% \title{Radio-transparent multi-layer insulation for large aperture receiver} 


% \author{}
%\thanks{}
% \affiliation{}

\author{J.~Choi}
\affiliation{Korea University, Anam-dong Seongbuk-gu, Seoul 136-713
Republic of Korea}
\author{H.~Ishitsuka}
\affiliation{Department of Particle and Nuclear Physics, School of High Energy Accelerator Science, The Graduate University for Advanced Studies (SOKENDAI), Shonan Village, Hayama, Kanagawa 240-0193 Japan}
\author{S.~Mima}
\affiliation{Terahertz Sensing and Imaging Team Terahertz-wave Research Group, RIKEN, 2-1 Hirosawa, Wako, Saitama 351-0198 Japan}
\author{S.~Oguri}
\email[]{shugo@post.kek.jp}
\affiliation{Institute of Particle and Nuclear Studies, High Energy Accelerator Research Organization~(KEK), Oho, Tsukuba, Ibaraki 305-0801 Japan}
\author{K.~Takahashi}
\affiliation{Department of Physics, Tohoku University, Sendai, Miyagi 980-8578 Japan}
\affiliation{Terahertz Sensing and Imaging Team Terahertz-wave Research Group, RIKEN, 2-1 Hirosawa, Wako, Saitama 351-0198 Japan}
\author{O.~Tajima}
\affiliation{Institute of Particle and Nuclear Studies, High Energy Accelerator Research Organization~(KEK), Oho, Tsukuba, Ibaraki 305-0801 Japan}
\affiliation{Department of Particle and Nuclear Physics, School of High Energy Accelerator Science, The Graduate University for Advanced Studies (SOKENDAI), Shonan Village, Hayama, Kanagawa 240-0193 Japan}


\date{\today}

\begin{abstract}
In the field of radiowave detection,
enlarging the receiver aperture
to enhance the amount of light detected
is essential for greater scientific achievements.
One challenge in using radio transmittable apertures
is keeping the detectors cool.
This is because transparency to thermal radiation
above the radio frequency range increases the thermal load.
In shielding from thermal radiation, 
a general strategy is to install thermal filters
in the light path between aperture and detectors.
However, there is difficulty
in fabricating metal mesh filters of large diameters.
It is also difficult to maintain large diameter absorptive-type filters
in cold because of their limited thermal conductance.
A technology that maintains cold conditions
while allowing larger apertures
has been long-awaited.
We propose radio-transparent multi-layer insulation (RT-MLI),
composed from a set of stacked insulating layers.
The insulator is transparent to radio frequencies,
but not transparent to infrared radiation.
The basic idea for cooling is similar to conventional multi-layer insulation.
It leads to a reduction in thermal radiation
while maintaining a uniform surface temperature.
The advantage of this technique over other filter types
is that no thermal links are required.
As insulator material, we used foamed polystyrene;
its low index of refraction
makes an anti-reflection coating unnecessary.
We measured the basic performance of RT-MLI
to confirm that thermal loads are lowered with more layers.
We also confirmed that our RT-MLI has high transmittance to radiowaves,
but blocks infrared radiation.
For example, RT-MLI with 12 layers has a transmittance
greater than 95\% (lower than 1\%) below 200~GHz (above 4~THz).
We demonstrated its effects in a system with absorptive-type filters,
where aperture diameters were 200 mm.
Low temperatures were successfully maintained for the filters.
We conclude that
this technology significantly enhances
the cooling of radiowave receivers,
and is particularly suitable for large-aperture systems.
This technology is expected to be applicable to various fields,
including radio astronomy, geo-environmental assessment,
and radar systems.
\end{abstract}


% \keywords{}

\maketitle %\maketitle must follow title, authors, abstract and \pacs

% If in two-column mode, this environment will change to single-column format so that long equations can be displayed. 
%\begin{widetext}
%$$\mbox{put long equation here}$$
%\end{widetext}

\section{Introduction}\label{intro}

Progress in the development of low-temperature detectors
has led to breakthroughs in various fields of radiowave detection
including radio astronomy~\cite{doi:10.1117/12.857871, Planck, Herschel}
and submillimeter-wave imaging
for geo-environmental assessment~\cite{SMILES}.
In particular, large detector arrays in a large-aperture system
enable faint signals to be detected
because the total amount of detected light is large.
This has led to greater scientific achievements in various fields,
such as cosmic microwave background radiation
measurements~\cite{doi:10.1117/12.926158}.
A major challenge in radiowave receiver systems
is keeping detectors cool.
Various techniques have been developed,
e.g., shading the inside of the aperture using absorptive-type filters
with dielectric materials~\cite{doi:10.1117/12.857871},
reflective-type filters (e.g., metal mesh filters~\cite{Ade_mmf}),
or combinations of these.
These filters are kept cold using thermal links
from the cold stages of a cryocooler or from a general cooling medium.

However, enlargement of the aperture leads to new difficulties,
such as in fabricating large-diameter metal mesh filters
and in maintaining uniformly cold temperature conditions
over absorptive-type filter surfaces.
Thermal re-emission from a warm filter then becomes a serious issue.
Therefore, a technology that enables large-aperture systems
to maintain a cold condition is required.

In this paper,
we propose radio-transparent multi-layer insulation~(RT-MLI)
composed of a set of stacked insulating layers.
The insulating material is transparent to radiowaves
but blocks infrared radiation.
The RT-MLI leads to a reduction in thermal loads
while maintaining a uniform surface temperature.
We measured the basic performance of RT-MLI
in terms of reduction in thermal load and transmittance.
We also demonstrated its cooling abilities
in combination with absorptive-type filters.

\section{Radio-transparent multi-layer insulation (RT-MLI)}

\subsection{Idea and design}

A multi-layer insulation~(MLI)%
~\cite{MLI1, MLI:lockheed_model, MLI:layer_by_layer_model, hedayat:1557}
is a thermal insulation system composed of multiple layers
of thin insulating sheets.
MLI is a major concept in thermal design
and is primarily intended to reduce the heat
contributed by thermal radiation. 
This technology is commonly used for applications in vacuum conditions,
where conduction and convection are significantly reduced
and heat radiation dominates. 
The principle behind MLI is to balance the thermal radiation between layers. 
More layers can be added to further reduce radiation losses.

%%%%%%%%%%%%%%%%%%%%%%%%%%%
\begin{table}[htb]
\caption{Physical properties of
 Styroace-II Styrofoam at room temperature%
 ~\footnote{http://www.dowkakoh.co.jp/styrofoam/data.html}.
\label{tab:styroace} 
}
\begin{tabular}{
	@{\hspace{0.2cm}}
	l
	@{\hspace{0.5cm}}
	l
	@{\hspace{0.2cm}}
	}
\hline
\hline
Material & Foamed polystyrene \\
Density & 0.025~g/cm$^3$ \\
Thermal conductivity & 0.028~W/m$\cdot$K \\
Coefficient of thermal expansion & $7 \times 10^{-5}$~K$^{-1}$ \\
Index of refraction & 1.03 \\
\hline
\hline
\end{tabular}
\end{table}
%%%%%%%%%%%%%%
\begin{figure}[htb]
\includegraphics[width=8.5cm]{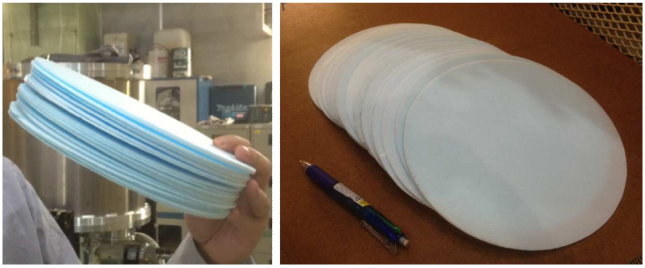}
\caption{Photos of an RT-MLI composed of stacked Styrofoam layers. 
 Each layer has a diameter of 210~mm, and a thickness of 3~mm.
 The total number of layers shown in the left photo is 17.
 \label{fig:pic_TMLI}
 }
\end{figure}
%%%%%%%%%%%%%%
Evaporated metal films are popular materials for MLI.
However, they are not transparent to radiowaves.
Replacement of evaporated metal films with a radio-transparent material
enables the same principle as MLI to be exploited for radiowaves.
Foamed polystyrene is an appropriate insulation material. 
It has high transmittance in the millimeter-wave range
but almost zero transmittance in the infrared region.
Additionally, anti-reflection coatings are unnecessary on the surface
because the index of refraction of foamed polystyrene is low.
We used a commercial material, Styroace-II Styrofoam,
provided by the Dow Chemical Company.
The physical properties of this material are summarized
in Table~\ref{tab:styroace}.
We fabricated a RT-MLI using a set of stacked Styrofoam layers,
as shown in Fig.~\ref{fig:pic_TMLI}.

\subsection{Principle}\label{sec:principle}
%%%%%%%%%%%%%%
\begin{figure}[htb]
\includegraphics[width=4.8cm]{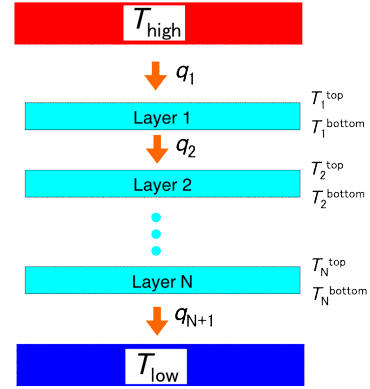}
\caption{Layout showing the principle behind RT-MLI.
 Exchanges of thermal radiation between the layers are balanced.
 The thermal loads conducted
 from the top surface to the bottom surface in each layer
 are also balanced with the exchanged radiation on each surface.
 \label{fig:principle}
 }
\end{figure}
%%%%%%%%%%%%%%
The principle underpinning RT-MLI is similar to that of conventional MLI.
As shown in Fig.~\ref{fig:principle},
thermal radiation exchanges between the layers are balanced.
By neglecting higher-order effects,
the exchanged thermal loads per unit area can be described as follows:
\begin{eqnarray}
	q_{1~~~~} &=&  \sigma \left[ \left( T_{\rm high} \right)^4 - \varepsilon \left( T_{1}^{\rm top} \right)^4 \right] ,
	\\
	q_{2~~~~} &=&  \sigma \left[ \varepsilon \left( T_{1}^{\rm bottom} \right)^4 - \varepsilon \left( T_{2}^{\rm top} \right)^4 \right] ,	
	\\
	& \vdots &	\nonumber
	\\		
	q_{N~~~} &=&  \sigma \left[ \varepsilon \left( T_{N-1}^{\rm bottom} \right)^4 - \varepsilon \left( T_{N}^{\rm top} \right)^4 \right] ,		
	\\		
	q_{N+1} &=&  \sigma \left[ \varepsilon \left( T_{N}^{\rm bottom} \right)^4 - \left( T_{\rm low} \right)^4 \right] ,		
\end{eqnarray}
where $q_{i}$, $\sigma$, and $\varepsilon$ denote
the load into the $i$-th layer from the direction of the previous layer,
the Stefan-Boltzmann constant,
and the effective emissivity of the Styrofoam, respectively.
Because each component of the radiation is balanced,
i.e., $q \equiv q_1 = q_2 = \cdots = q_{N+1}$,
summation of the above equations over all layers gives:
\begin{eqnarray}
	q &=&  \frac{1}{N+1}\cdot \sigma \left[ \left( T_{\rm high} \right)^4 - \left( T_{\rm low} \right)^4 \right] \nonumber
	\\
	&&
		- \frac{1}{N+1}\cdot \varepsilon \sigma \sum_{i=1}^{N} 	\left[ \left( T_{i}^{\rm top} \right)^4 - \left( T_{i}^{\rm bottom} \right)^4 \right] . \label{eq:2ndterm}
\end{eqnarray}

Inside each layer,
the thermal conductance is also balanced
with respect to the thermal radiation for each layer surface:
\begin{eqnarray}
	q &=& \frac{\kappa}{d} 
		\left( 
		T_i^{\rm top} - T_i^{\rm bottom}
		\right) , \label{eq:conductance}
\end{eqnarray}
where $\kappa$ and $d$ are
the thermal conductivity and layer thickness, respectively.
This equation indicates
that the contribution of the second term in Eq.~\ref{eq:2ndterm}
is significant when the magnitude of $q$ is large
(i.e., the number of layers is small)~\footnote{
The second term in Eq.~\ref{eq:2ndterm} is also significant
when the difference between
the temperatures $T_{\rm high}$ and $T_{\rm low}$ is small
(this is true in general when $T_{\rm high}$ is small).
However,
the effects of stray light is more problematic in a real system;
the absorption of stray light can degrade the performance of an RT-MLI
when $T_{\rm high}$ is small.}.

For conventional MLI,
the second term in Eq.~\ref{eq:2ndterm} is negligible
because of $T_i^{\rm top}  = T_i^{\rm bottom}$.
With this approximation, Eq.~\ref{eq:2ndterm} reduces to:
\begin{eqnarray}
	q &=&  \frac{1}{N+1} \cdot \sigma \left[ \left( T_{\rm high} \right)^4 - \left( T_{\rm low} \right)^4 \right]. \label{eq:1/(N+1)-law}
\end{eqnarray}
The thermal radiation is proportional to
the inverse of the number of layers plus one, i.e., $q \propto 1/(N+1)$.
This simplified model for the conventional MLI
is called the $1/(N+1)$ law in this paper.
For RT-MLI, the balanced radiation tends to be smaller
than that predicted by the $1/(N+1)$ law
because $T_i^{\rm top}  > T_i^{\rm bottom}$.
Therefore, a dedicated simulation has to be performed
by simultaneously solving Eqs.~\ref{eq:2ndterm} and \ref{eq:conductance}.

Another unique feature of RT-MLI
is that the above equations are independent of the layer area.
This means that the uniformity of the surface temperature
is guaranteed in principle.
RT-MLI is applicable to large-aperture systems
without any change in thermal shielding performance;
this is a major advantage of the technology.

\subsection{Reduction of thermal radiation}

\begin{figure}[htb]
\includegraphics[width=7.4cm]{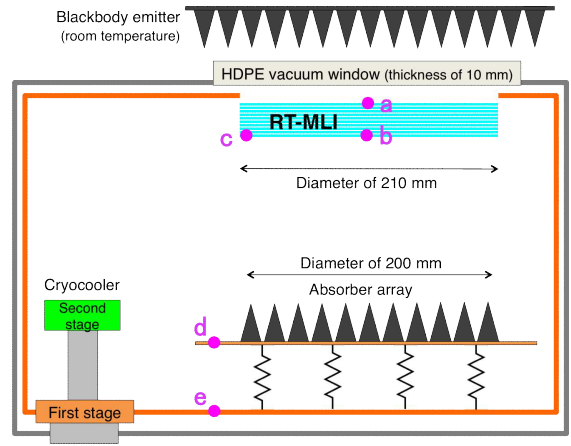}
\caption{Layout for measuring the basic performance of the RT-MLI
 in reducing room temperature radiation.
 Five thermometers were located at the positions marked by solid dots, a--e.
 We monitored the top and bottom surface temperatures of the RT-MLI
 as well as the temperatures on the absorber array and the radiation shield.
 Thermal loads passing through the RT-MLI were measured
 by obtaining the temperature difference between thermometers d and e.
 \label{fig:layout_of_setup_nlayers}
 }
\end{figure}

The performance in terms of the reduction in thermal radiation was
measured using the setup shown in Fig.~\ref{fig:layout_of_setup_nlayers}.
The cryostat consisted of a vacuum chamber and
a copper radiation shield inside the chamber.
A two-stage Gifford-McMahon (GM) cryocooler
(Sumitomo Heavy Industries Ltd., RDK-408S)
maintained cold conditions inside the cryostat;
the first stage maintained shield temperatures at around 27~K,
and no thermal link from the second stage was used in this test.
The cylindrically shaped chamber had a diameter of 508~mm
and height of 480~mm.
There was a circular aperture of diameter 260~mm on top of the chamber.
This aperture was sealed with a high-density polyethylene window.
The aperture of the shield was 210~mm in diameter.
A pyramid-shaped absorber array was located under the aperture.
Each absorber piece consisted of an Eccosorb-coated aluminum block
and therefore had good thermal conductance,
similar to that of a metal~\cite{kogut:5079, hasegawa:054501}.
The absorbers were set on a copper plate (of thickness 5~mm)
that was also cooled by the first stage of the cryocooler;
this resulted in a uniform temperature
across all absorbers~\cite{hasegawa:054501}.
The thermal conductance between the copper plate and the cryocooler
was approximately 1~W/K.
A blackbody emitter
(Eccosorb CV-3 from Emerson \& Cuming Microwave Products, Inc.)
at room temperature ($\sim$298~K) was also set outside the vacuum window.

%%%%%%%%%%%%%%
%%%%%%%%%%%%%%

We set the RT-MLI layers behind the shield aperture
using a Styrofoam cylinder
(inner and outer diameters 215~mm and 225~mm, respectively,
and height 50~mm).
At the bottom of the cylinder,
small tabs prevented the RT-MLI layers from falling.
The RT-MLI layers were placed in the cylinder without any pressure
except that due to gravity.
The cylinder was placed on top of Styrofoam columns
that were placed at the edge of the copper plate of the absorbers.
The upper side of the RT-MLI directly faced the vacuum window.
The thermal conductivity of the support structure was negligible;
the RT-MLI did not indicate any cooling other than radiative cooling.
While varying the number of Styrofoam layers used,
we measured the temperatures achieved at the locations
shown in Fig.~\ref{fig:layout_of_setup_nlayers}.

%%%%%%%%%%%%%%
\begin{table}[htb]
\caption{Temperatures (K) achieved at each thermometer location
 (see Fig.~\ref{fig:layout_of_setup_nlayers}).
 \label{tab:nlayers} 
 }
\begin{tabular}{
	@{\hspace{0.1cm}}
	l
	@{\hspace{0.3cm}}
	r
	@{\hspace{0.3cm}}
	r
	@{\hspace{0.3cm}}
	r	
	@{\hspace{0.3cm}}
	r		
	@{\hspace{0.1cm}}	
	}
\hline
\hline
 & \multicolumn{4}{c}{RT-MLI: number of layers} \\
& \multicolumn{1}{c}{3} & \multicolumn{1}{c}{6} & \multicolumn{1}{c}{12} & \multicolumn{1}{c}{17} \\
\hline
a: RT-MLI (top-center) & 259.4 & 267.4 & 274.5 &  282.4 \\
b: RT-MLI (bottom-center) & 186.5 & 171.8 & 154.7 &  142.5 \\
c: RT-MLI (bottom-edge) & 185.4 & 169.3 & 154.4 &  142.1  \\
d: Absorber             &   29.9 &  29.3  &   28.6 &  28.1  \\
e: GM first stage&   27.6 &  27.4  &    27.2 & 27.0  \\
\hline
\hline
\end{tabular}
\end{table}
%%%%%%%%%%%%%%
%%%%%%%%%%%%%%
\begin{figure}[htb]
\includegraphics[width=8.4cm]{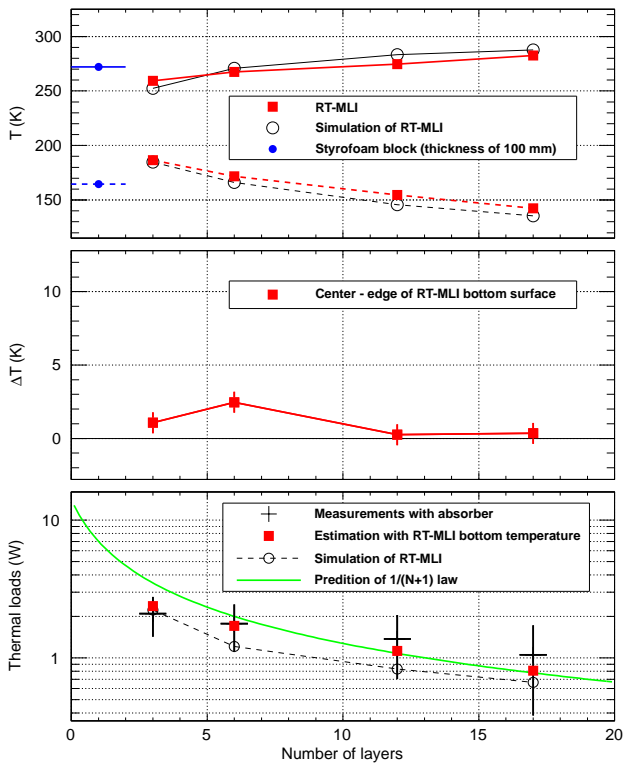}
\caption{[Top panel]: Surface temperatures of RT-MLI
 as a function of the number of Styrofoam layers;
 the solid and dashed lines indicate the top side
 (location a in Fig.~\ref{fig:layout_of_setup_nlayers})
 and the bottom side
 (location b in Fig.~\ref{fig:layout_of_setup_nlayers})
 of the RT-MLI, respectively.
 Surface temperatures of a thick Styrofoam block (thickness of 100~mm)
 are also shown for comparison.
 Several layers of the RT-MLI with a total thickness of $\sim$20~mm
 achieved a performance equivalent to that of the thick Styrofoam block.
 A simulation with an emissivity of 0.72 almost exactly reproduces
 the temperature gradient of the RT-MLI.
 [Middle panel]: Temperature difference between the two thermometers
 on the bottom surface of the RT-MLI in the center
 (b in Fig.~\ref{fig:layout_of_setup_nlayers})
 and on the edge
 (c in Fig.~\ref{fig:layout_of_setup_nlayers}).
 The principle of the RT-MLI guarantees uniform surface temperatures
 (see main text for details: section \ref{sec:principle}).
 [Bottom panel]:
 Thermal loads passing into the absorber array for each configuration.
 The thermal radiation estimated
 from the bottom temperature of the RT-MLI is also shown;
 it should be the dominant source of the thermal loads.
 A prediction based on the simple $1/(N+1)$ law
 (Eq.~\ref{eq:1/(N+1)-law}),
 regarding the behavior of a conventional MLI
 is overlaid for comparison.
 The RT-MLI exhibits a lower slope than the simple $1/(N+1)$ law
 because of a contribution of the second term in Eq.~\ref{eq:2ndterm}.
 \label{fig:nlayers_dependence}
 }
\end{figure}
%%%%%%%%%%%%%%

The temperatures for each configuration are summarized
in Table~\ref{tab:nlayers}.
The bottom side of the RT-MLI achieved lower temperatures
as the number of layers increased.
In contrast, the top side temperature increased
because of the smaller temperature gradient
associated with the larger number of layers,
as predicted by Eq.~\ref{eq:2ndterm}.
A comparison between the temperatures obtained
from measurements and those from simulations
based on Eq.~\ref{eq:2ndterm} and Eq.~\ref{eq:conductance}
is shown in the top panel of Fig.~\ref{fig:nlayers_dependence}.
The middle panel of Fig.~\ref{fig:nlayers_dependence}
demonstrates temperature uniformity across the RT-MLI surface.
For each configuration,
the thermal loads passing into the absorbers
are shown in the bottom panel of Fig.~\ref{fig:nlayers_dependence}. 
Estimation of the thermal radiation from the bottom surface temperature
of the RT-MLI is also shown;
it should be the dominant source of thermal load.
The reduction in thermal load was roughly verified
using a typical heat capacity curve of the cryocooler~\footnote{
A typical heat capacity curve is found on the website of the company:
http:\slash\slash{}www.janis.com\slash{}Libraries\slash{}4K\_Coldheads\slash{}RDK-408S\_cryocooler\_typical\_load\_map.sflb.ashx}; 
in shifting from a 3-layer to a 17-layer configuration,
the reduction in thermal load was approximately 1 W.
This confirmed the reduction in thermal load,
which follows the description provided in section \ref{sec:principle}.

\subsection{Transmittance}

The transmittance of the RT-MLI was measured
using two different systems:
a Fourier transform spectrometer~(FTS)~\cite{FTS},
and a radiowave signal generator.
The FTS system consisted of Fourier transform infrared spectrometers
from the Japan Spectroscopic Corporation
(JASCO, Tokyo, Japan)
having a combination of two sensors:
an indium antimonide (InSb) hot electron bolometer
(QMC Instruments Ltd., Cardiff, United Kingdom)
for measurements between 180~GHz and 1.6~THz,
and a pyroelectric detector for measurements above 1.6~THz.
In the latter system,
the radiowave signals were generated by a signal generator
(E8247C, Agilent Technologies Inc., Santa Clara, California)
with multiplication performed using the AMC-10-RFH00 (AMC-05-RFH00)
of Millitech Inc. (Northampton, Massachusetts)
for the 80~GHz--110~GHz (140~GHz--220~GHz) frequency band.
The intensity of the signal was measured using diode detectors:
DET-10 (or DET-05) of Millitech Inc.
for the 80~GHz--110~GHz (or 140~GHz--220~GHz) frequency band.
The transmittance was obtained from the ratio
of the measured signal intensity before inserting the RT-MLI
to that after its insertion.

%%%%%%%%%%%%%%
\begin{figure}[htb]
\includegraphics[width=8.5cm]{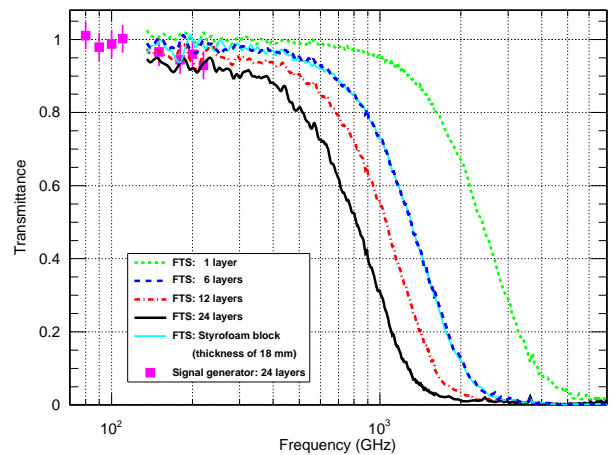}
\caption{Transmittance of RT-MLI at room temperature.
 Using an FTS,
 we measured the transmittances of four different configurations
 distinguished by the number of layers 1, 6, 12, and 24.
 We also measured the transmittance of a Styrofoam block for comparison.
 Below the 220~GHz region, the transmittance of the 24-layer sample
 was also measured using signal generators.
 The transmittance was roughly proportional
 to the $N$-th power of the number of layers,
 i.e., the transmittance of an $N$-layer RT-MLI was
 approximately $0.997^N$ at a frequency of 200~GHz.
 \label{fig:transmittance}
}
\end{figure}
%%%%%%%%%%%%%%
The measured transmittance for each layer configuration
is shown in Fig.~\ref{fig:transmittance}.
We did not deposit any anti-reflection coating on the RT-MLI surface.
The logarithm of the transmittance was roughly proportional
to the number of layers ($N$),
i.e., the transmittance of an $N$-layer RT-MLI is
greater than $0.997^N$ below a frequency of 200~GHz.
We confirmed the high transmittance of the RT-MLI at radiowave frequencies.
The transmittance of a Styrofoam block
is also shown in Fig.~\ref{fig:transmittance}.
The Styrofoam block thickness (18~mm)
was equivalent to the thickness of the 6-layer RT-MLI;
a consistent transmittance was obtained.

\section{Demonstration with a combination of RT-MLI and absorptive-type filters}

\begin{figure}[htb]
\includegraphics[width=7.2cm]{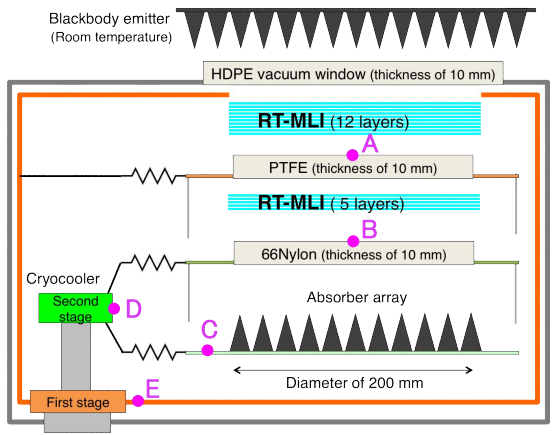}
\caption{Layout of the setup used to demonstrate the effects of RT-MLI.
 This setup emulates a radiowave receiver with a 200-mm-diameter aperture.
 Instead of a detector array, we used an absorber array.
 The temperatures obtained at each location (A, B, $\cdots$, E) are summarized in Table~\ref{tab:phi200}.
 \label{fig:layout_of_setup_phi200}
 }
\end{figure}

\begin{figure}[htb]
\includegraphics[width=7.3cm]{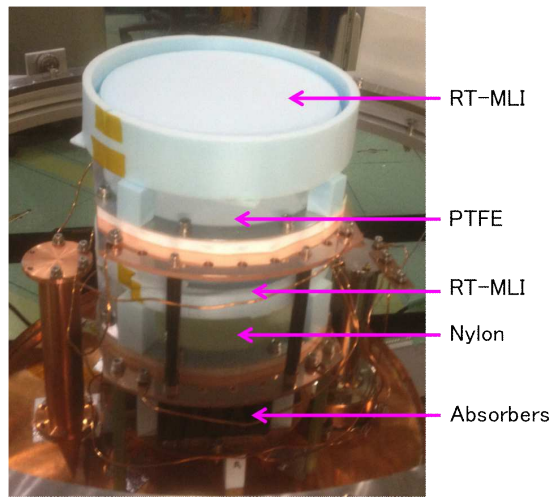}
\caption{Photograph of the setup shown in Fig.~\ref{fig:layout_of_setup_phi200}.
 \label{fig:photo_setup_phi200}
 }
\end{figure}

We also demonstrated the effects of RT-MLI
in a system having absorptive-type filters.
We emulated the receiver system
shown in Fig.~\ref{fig:layout_of_setup_phi200}
and Fig.~\ref{fig:photo_setup_phi200},
but we used an absorber array instead of a detector array.
We installed two sets of RT-MLIs between each layer,
including the layers of the vacuum window,
polytetrafluoroethylene (PTFE) filter, and 66-Nylon filter.
Using copper wires (2~mm diameter) and copper jigs
that held the filters,
the PTFE filter was thermally linked
to the first stage of the GM cryocooler.
The Nylon filter and the absorbers were also linked
to the second stage of the GM cryocooler. 
For the demonstration,
the thermal conductance values of each thermal link
were maintained at the low value of approximately 0.05~W/K.
We did not need any thermal link to the RT-MLI
because the heat was expelled through radiative cooling.
Therefore, the installation of the RT-MLI
did not place any additional load on the cryocooler.

%%%%%%%%%%%%%%
%%%%%%%%%%%%%%
%%%%%%%%%%%%%%
%%%%%%%%%%%%%%

We monitored the temperature at the center of each filter,
the edges of the absorber and Nylon filter,
and both cold heads of the cryocooler.
The temperatures obtained before and after the installation
of the RT-MLI are summarized in Table~\ref{tab:phi200}.
Also shown are the temperatures measured from the installation
of two Styrofoam blocks rather than the RT-MLI layers.
The thicknesses of the layers and the blocks were equivalent,
i.e., we replaced 12 and 5 layers of RT-MLI
with Styrofoam blocks of thicknesses 40~mm and 15~mm, respectively.
RT-MLI achieved better performance than the Styrofoam block.
Because of the low thermal conductivity values
($\sim$0.25~W/m$\cdot$K) of PTFE and Nylon,
the temperatures of both filters were very high
before the installation of RT-MLI.
After installation of the RT-MLI layers,
we confirmed significant improvements
in the temperatures obtained at each location.
The reduction in thermal loads also resulted in
a small temperature gradient across the filters;
e.g., the temperature difference between the center and edge
of the Nylon filter
was significantly reduced from 85.0~K (no additional shielding)
to 11.4~K (after the installation of RT-MLI).

\begin{table}[htb]
\caption{Temperatures (K) obtained the setup
 in Fig.~\ref{fig:layout_of_setup_phi200};
 the additional radiation shielding took
 the form of no shielding,
 installation of the Styrofoam blocks,
 and installation of RT-MLI
 (12~layers above the PTFE filter and 5~layers above the Nylon filter).
 \label{tab:phi200} 
 }
\begin{tabular}{
	@{\hspace{0.1cm}}
	l
	@{\hspace{0.3cm}}
	r
	@{\hspace{0.3cm}}
	r	
	@{\hspace{0.3cm}}
	r		
	@{\hspace{0.1cm}}	
	}
\hline
\hline
Additional shields & Nothing & Styrofoam block & RT-MLI \\
\hline
A : PTFE                     & 218.9   & 133.8  &    95.7 \\
B : Nylon (center)        & 128.5   &   48.7  &    36.3 \\
B': Nylon (edge)          &   43.5   &   31.5  &    24.9 \\
C : Absorber                &   15.5   &     9.2  &     8.2 \\
D : GM second stage  &     8.0    &     6.2  &     5.9 \\
E : GM ~first~~ stage  &   29.6   &   27.1  &   26.7 \\
\hline
\hline
\end{tabular}
\end{table}
%%%%%%%%%%%%%%%%%%%%%%%%%%%%%%%%%%%%%

\begin{figure}[htb]
\includegraphics[width=7.cm]{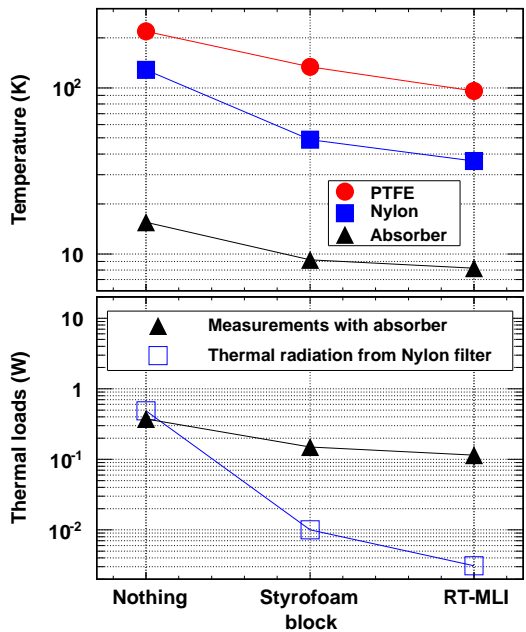}
\caption{[Top]: Temperatures attained at each layer as shown
 in Fig.~\ref{fig:layout_of_setup_phi200} and Table~\ref{tab:phi200}.
 [Bottom]: Thermal loads passing into absorbers measured
 using temperature differences between C and D
 in Fig.~\ref{fig:layout_of_setup_phi200} and Table~\ref{tab:phi200}.
 Estimated thermal radiation from the Nylon filter is also shown.
 The difference between the measured load and the estimated radiation
 ($\approx 0.1$~W)
 indicates that effects of reflected radiations from the upper layers.
 A setup that has tight shielding to prevent reflection
 is expected to reduce the difference.
 \label{fig:loads_phi200}
 }
\end{figure}

We confirmed a significant reduction in the thermal loads
passing into the absorbers (Fig.~\ref{fig:loads_phi200}).
The lower temperatures achieved in the cryocooler also indicated
a reduction in thermal load of the system.
This cryocooler achieved temperatures of 26.5~K and 5.5~K
for each stage without any additional load from outside the radiation shield.
The typical heat capacity curve of the cryocooler indicates
a significant load reduction of $\sim 4$~W ($\sim 3$~W)
for the first (second) stage.
We confirmed that RT-MLI is useful
to shield the aperture from outside thermal radiation.

%%%%%%%%%%%%%%
%%%%%%%%%%%%%%

\section{Conclusion}
We developed RT-MLI composed of a set of stacked layers of polystyrene foam.
RT-MLI reduces thermal radiation while maintaining transparency to radiowaves.
We confirmed both surface temperature uniformity
and a reduction in thermal loads.
No anti-reflection coating is necessary.
We also confirmed that RT-MLI has high transmittance
for radiowaves but blocks infrared radiation.
Another advantage of the RT-MLI compared with other filter types
is that no thermal links are necessary.
We demonstrated the effects of RT-MLI in combination
with absorptive-type filters in a large-aperture setup
(where the aperture diameter of each filter was 200~mm);
low filter temperatures were successfully maintained.
The principle of RT-MLI also guarantees
application to significantly larger aperture systems,
e.g., those with diameters of a few meters.

We conclude that RT-MLI is a useful technology
in the cooling of radiowave receiver systems
and particularly for the enlargement of a system aperture
while maintaining a uniform surface temperature
and high-transmittance conditions.
RT-MLI can possibly be applied in various fields
including radio astronomy, geo-environmental assessment,
and radio detection and ranging (radar) systems.

\begin{acknowledgments}
This work was supported by Grants-in-Aid for Scientific Research from the Ministry of Education, Culture, Sports, Science and Technology, Japan (KAKENHI 23684017, 21111003, and 25610064).
It was also supported by FY 2012 Joint Development Research on an Open Application Basis Program of the National Astronomical Observatory of Japan (NAOJ),
Research Grants in the Natural Sciences from the Mitsubishi Foundation,
and the Basic Science Research Program through the National Research Foundation of Korea (NRF) funded by the Ministry of Education, Science and Technology (2013R1A1A2004972).
We thank Masanori Kawai and Chiko Otani for discussions with regard to cryogenics and the applications of this work.
We also thank Takayuki Tomaru and Mitsuhiro Yoshida, who kindly allowed us to use instruments for the transmittance measurements.
We thank Masaya Hasegawa and Masashi Hazumi for providing a styrene foam cutter, which was a very useful tool for the tests performed in this paper. 
Finally, we thank Ken'ichi Karatsu, Akito Kusaka, Tomotake Matsumura, Yuji Chinone, and Masato Naruse, with whom we had useful discussions on the design concepts of the radiowave receivers.
\end{acknowledgments}

%\bibliography{ref.bib}
% Run this once to generate your BBL file. Then copy the contents of your BBL file into your main latex file, commenting out "\bibliography"

\end{document}